\documentclass[aps,twocolumn,showpacs]{revtex4}

\usepackage[xdvi]{graphicx}
\usepackage{latexsym}
\usepackage{amsbsy}
\usepackage{amsmath}
\usepackage{amssymb}

\begin{document}

\date{\today}

\title{Quantum spin dynamics}

\author{R.\ Wieser}
\affiliation{Institut f\"ur Angewandte Physik und Zentrum f\"ur
  Mikrostrukturforschung, Universit\"at Hamburg, Jungiusstrasse 11, D-20355
  Hamburg, Germany}

\begin{abstract}
 The classical Landau-Lifshitz equation has been derived from quantum
 mechanics. Starting point is the assumption of a non-Hermitian
 Hamilton operator to take the energy dissipation into account. The
 corresponding quantum mechanical time dependent Schr\"odinger,
 Liouville and Heisenberg equation have been described and the
 similarities and differences between classical and quantum mechanical
 spin dynamics have been discussed. Furthermore, a time dependent
 Schr\"odinger equation corresponding to the classical
 Landau-Lifshitz-Gilbert equation and two ways to include temperature
 into the quantum mechanical spin dynamics have been proposed.     
\end{abstract}

\pacs{75.78.-n, 75.10.Jm, 75.10.Hk}
\maketitle

\section{INTRODUCTION} \label{s:intro}
The increasing need of faster and more powerful computer technology
provides more and more new fields in magnetism like the field of magnonics
\cite{kruglyakJPD10} or the field antiferromagnetic spintronics
\cite{wieserPRL08,wieserPRL11,macdonaldPTRS11,barthemNatCommun13}. In
all these new fields as well as in the 
established areas like the ferromagnetic spintronics
\cite{parkinSCIENCE08,schiebackEPJB07,ralphJMMM08,wieserPRB10ii} the
interest is mostly focused in the 
dynamics. To describe the dynamics of magnetic structures it is need
to have an equation of motion. In the most cases this equation is the
Landau-Lifshitz-Gilbert (LLG) equation \cite{GilbertIEEE04}. Together 
with the Maxwell equations \cite{jacksonBOOK99} this equation is the main
equation of Micromagnetism \cite{brownBOOK63}, the field which
describes the dynamics of nearly all magnetic devices of our daily
life. There are more equations describing the dynamics in magnetism
like the Bloch equation \cite{blochPR46}, the Landau-Lifshitz-Bloch equation
\cite{garaninPA91,garaninPRB97,evansPRB12} 
or the Ishimori equation \cite{ishimoriPTP84}. However, the  most
important equation is the LLG equation. 

The LLG equation was originally introduced by T. L. Gilbert in 1955
\cite{GilbertBOOK55} as a reaction of the fact that the
Landau-Lifshitz (LL) equation which was proposed by L. D. Landau and
E. M. Lifshitz in 1932 contains a purely 
phenomenological damping term \cite{landauPZS32}. Gilbert's original 
equation, the Gilbert equation, is an implicit differential
equation. The explicit form of the Gilbert equation is similar to the
original LL equation, therefore this equation is called
Landau-Lifshitz-Gilbert (LLG) equation. Both equations contain two
terms describing the dynamics of a single spin. The first term
describes a precessional motion within an  
effective field and the second term a relaxation of the spin into the
direction of the effective field. During this relaxation the energy
will be reduced. The difference between the LL and LLG equation is the
fact that in the LLG equation both terms, the precessional as well
as the relaxation term are damped. In the original LL equation the
precession term does not contain a damping constant, which means that
this motion is not affected by the damping.   

In 1956 Kikuchi \cite{kikuchiJAP56} has shown that the LLG
equation describes the correct physics while the LL equation fails in
the limit of a huge damping. The reason is the missing damping for the
precession in the LL equation. 

The goal of this publication is to show that the LLG equation can be
derived from quantum mechanics which gives new insite in the
underlying physics and makes it possible to study the
quantum-classical transition. Text books
\cite{fogedbyBOOK80,guoBOOK08} as well as previous publications
\cite{lakshmananPTRS11,sakumaArxiv06,wieserPRB11} 
describe only the derivation of the precessional term. In all cases,
the damping term has been added later phenomenological. However,
complete derivations using a classical description are known
\cite{GilbertIEEE04,fahnleJPD08}. 

Furthermore, the paper discusses the difference between the classical
and quantum mechanical description of the spin dynamics and shall
correct the mistakes made in the former publication \cite{wieserPRL13}.

The manuscript is organized as follows. After this introduction, in 
Sec. \ref{s:LLG}, the equation of motion for a single spin will be
derived and extended to a multi-spin system
(Sec. \ref{s:oneormore}). In Sec. \ref{s:simulations} the
results of the previous sections will be proved by numerical
calculations and the last Sec. \ref{s:Temperature} describes two ways
how to include temperature and quantum fluctuations into the computer
simulations. The publication ends with a summary (Sec. \ref{s:summary}).

\section{EQUATION OF MOTION} \label{s:LLG} 
It is a well know fact that a non-Hermitian Hamiltonian $\hat{\cal H}
= \hat{\mathrm{H}} - i \lambda \hat{\Gamma}$, with $\hat{\mathrm{H}}$
and $\hat{\Gamma}$ Hermitian operators and $\lambda \in
\mathbb{R}^+_0$ a constant, lead to energy dissipation
\cite{weisskopfZP30,liboffPD04,kosikPHDThesis}. On the other hand such
a Hamiltonian does not conserve the norm of the wave function $|\psi
(t) \rangle = \exp(-i\hat{\cal H}t)|\psi_0 \rangle$:
\begin{equation}
n = \langle \psi(t) |\psi(t) \rangle = \langle \psi_0 |
e^{-2\lambda\hat{\Gamma}t} |\psi_0 \rangle = e^{-2\lambda\langle
    \hat{\Gamma} \rangle t} \;.   
\end{equation}
However, the norm can be conserved by replacing $\hat{\Gamma}$ by
$\hat{\Gamma} - \langle \hat{\Gamma} \rangle$. In this case we the
following wave function: 
\begin{equation} \label{PsiTime}
|\psi (t) \rangle = e^{-i\hat{\mathrm H}t}
e^{-\lambda\hat{\Gamma}t}e^{\lambda \langle \hat{\Gamma}
  \rangle t} |\psi_0 
\rangle \;,
\end{equation}
which is conserved: $n = 1$. The corresponding Schr\"odinger equation
is given by: 
\begin{eqnarray}
i\frac{\mathrm{d}}{\mathrm{d}t}|\psi(t)\rangle =
(\hat{\mathrm{H}} - i \lambda [\hat{\Gamma} - 
  \langle \hat{\Gamma} \rangle ])|\psi(t)\rangle \;.
\end{eqnarray}
The energy dissipation itself depends on $\lambda$ and
$\hat{\Gamma}$. In the following we assume that $\hat{\Gamma}$ is equal
$\hat{\mathrm{H}}$ and therefore, $\hat{\cal{H}} = \hat{\mathrm{H}} -
i\lambda(\hat{\mathrm{H}} - \langle \hat{\mathrm{H}} \rangle)$. With
this assumption the Schr\"odinger equation becomes \cite{gisinHelvPhysActa81}:
\begin{eqnarray} \label{TDSE_LL}
i\frac{\mathrm{d}}{\mathrm{d}t}|\psi(t)\rangle =
(\hat{\mathrm{H}} - i \lambda [\hat{\mathrm{H}} - 
  \langle \hat{\mathrm{H}} \rangle ])|\psi(t)\rangle \;.
\end{eqnarray}
We will see that this Schr\"odinger equation can be seen as the
quantum mechanical analog of the classical Landau-Lifshitz
equation, where $\lambda$ is the damping constant. 

Now, it is quite easy to show that this equation corresponds to the
following Liouville (von Neumann) equation \cite{wieserPRL13}:
\begin{equation} \label{Liouville}
\frac{\mathrm{d}\hat{\rho}}{\mathrm{d}t} =
i[\hat{\rho},\hat{\mathrm{H}}] -
  \lambda[\hat{\rho},[\hat{\rho},\hat{\mathrm{H}}]] \;.
\end{equation}
We assume that the system is in a pure state therefore the density
operator is given by:
\begin{equation} \label{PureDen}
\hat{\rho} = 
|\psi(t)\rangle\langle \psi(t) | \;.
\end{equation} 
Equation (\ref{Liouville}) has already the form of the
Landau-Lifshitz equation because the commutator $[\;,\;]$ in 
the case of spin systems lead to a vector product.  

So far, we have derived the quantum mechanical Liouville equation. The 
next step is the Heisenberg equation. Therefore, we concentrate
ourself first on the time dependence of the expectation value $\langle 
\hat{\mathbf{S}} \rangle$:
\begin{equation} \label{STime}
\frac{\mathrm{d}}{\mathrm{d}t}\langle \hat{\mathbf{S}} \rangle =
\mathrm{Tr}\!\left(\frac{\mathrm{d}}{\mathrm{d}t}\hat{\rho}
\hat{\mathbf{S}} \right) \;.
\end{equation} 
Including the Liouville equation (\ref{Liouville}) in
Eq.~(\ref{STime}), written in the alternative form:
\begin{equation}
\frac{\mathrm{d}\hat{\rho}}{\mathrm{d}t} =
i[\hat{\rho},\hat{\mathrm{H}}] -
  \lambda\left([\hat{\rho},\hat{\mathrm{H}}]_+ -
  2\hat{\rho}\hat{\mathrm{H}}\hat{\rho}\right) \;, 
\end{equation}
and get immediately the following differential
equation:
\begin{equation} \label{STime2}
\frac{\mathrm{d}}{\mathrm{d}t}\langle \hat{\mathbf{S}} \rangle =
-i \langle [\hat{\mathbf{S}},\hat{\mathrm{H}}] \rangle -
\lambda\left(\langle
     [\hat{\mathbf{S}},\hat{\mathrm{H}}]_+ \rangle - 2 \langle
     \hat{\mathrm{H}} \rangle\langle \hat{\mathbf{S}} \rangle \right) \;. 
\end{equation}
Here, the $[\;,\;]_+$ is the anticommutator and we have used the fact
that the trace is invariant under cyclic permutations. Furthermore:
\begin{eqnarray}
\mathrm{Tr}\!\left(\hat{\rho}\hat{\mathrm{H}}\hat{\rho}\hat{\mathbf{S}}\right)
&=& \sum\limits_m \langle m | \psi \rangle \langle \psi
|\hat{\mathrm{H}} | \psi \rangle \langle \psi | \hat{\mathbf{S}} | \psi
\rangle \langle \psi | m \rangle \nonumber \\
&=& \sum\limits_m \langle \psi |\hat{\mathrm{H}} | \psi \rangle
\langle \psi | \hat{\mathbf{S}} | \psi \rangle \langle \psi | m \rangle
\langle m | \psi \rangle \nonumber \\
&=& \langle \psi |\hat{\mathrm{H}} | \psi \rangle
\langle \psi | \hat{\mathbf{S}} | \psi \rangle = \langle
\hat{\mathrm{H}} \rangle \langle  \hat{\mathbf{S}}  \rangle \nonumber \\
&=& \mathrm{Tr}\!\left(\hat{\rho}\hat{\mathrm{H}}\right)
\mathrm{Tr}\!\left(\hat{\rho} \hat{\mathbf{S}}\right)  \;.
\end{eqnarray}   
Then, the Heisenberg equation appears after replacing the expectation values
$\langle \hat{\mathbf{S}} \rangle$ by $\hat{\mathbf{S}}$:
\begin{equation} \label{Heisenberg}
\frac{\mathrm{d}}{\mathrm{d}t} \hat{\mathbf{S}}  =
-i [\hat{\mathbf{S}},\hat{\mathrm{H}}] -
\lambda\left([\hat{\mathbf{S}},\hat{\mathrm{H}}]_+ - 2 \langle
     \hat{\mathrm{H}} \rangle \hat{\mathbf{S}} \right) \;. 
\end{equation} 
Now, the Heisenberg equation (\ref{Heisenberg}) as well as the
Liouville equation (\ref{Liouville}) can be used to derive the
Landau-Lifshitz equation. Therefore, we have to insert these
equations in (\ref{STime}) and to determine the traces. However,
we need to know the exact form of the density operator $\hat{\rho}$. 

In the case of a Heisenberg spin system with spin quantum number $S$
the density operator is given by the following multivector expansion
\cite{hofmannPRA04}: 
\begin{equation}
\hat{\rho} =
\frac{1}{2S+1}\hat{\mathbf{1}} +
\frac{1}{n_S}\sum\limits_{m}\langle
\hat{S}_m \rangle \hat{S}_m + \frac{1}{n_{2S}}\sum\limits_{ml}\langle
\hat{S}_{ml} \rangle \hat{S}_{ml} + \ldots 
\end{equation}
The first term is the Identity matrix which behaves under rotation
like a scalar. The next term is the sum over the three spin matrices
$\hat{S}_x$, $\hat{S}_y$, and $\hat{S}_z$. These matrices behave under
rotation like a vector. The next term is the sum 
over bivectors 
\begin{equation} \label{Bivector}
\hat{S}_{ml} = \frac{1}{2}[\hat{S}_m,\hat{S}_l]_+ -
\frac{1}{3}S(S+1)\delta_{ml} \;,
\end{equation}
and the next term (not shown) is the sum over
trivectors $\hat{S}_{mlk}$ and so on. All higher order terms have the
same scheme as the previous ones but with higher order tensors. The
prefactors $n_S$ and $n_{2S}$ are given by traces
$n_S = \mathrm{Tr}(\hat{S}_a\hat{S}_a)$ and $n_{2S} =
\mathrm{Tr}(\hat{S}_{ab}\hat{S}_{ab})$ (these values only depend on
the spin quantum number $S$ and are independent of the direction $a,b \in
\{x,y,z\}$). The prefactors of the higher expansion terms are similar. 

This expansion is highly related to the magnetic multipol expansion
\cite{vedmedenkoCPC08},  
and the number of terms depend on the spin quantum number $S$. In the
case $S = 1/2$ the expansion ends with the vector term 
\begin{equation} \label{RhoEinHalb}
\hat{\rho} =
\frac{1}{2}\left(\hat{\mathbf{1}} + \sum\limits_{m}\langle
\hat{\sigma}_m \rangle \hat{\sigma}_m\right) \;.
\end{equation}
Here, $\hat{\sigma}_m$ are the Pauli matrices. In the case $S=1$ the
expansion ends with the bivector term  
\begin{equation} \label{RhoEins}
\hat{\rho} =
\frac{1}{3}\hat{\mathbf{1}} + \frac{1}{2}\sum\limits_{m}\langle
\hat{S}_m \rangle \hat{S}_m + \frac{1}{2}\sum\limits_{ml}\langle
\hat{S}_{ml} \rangle \hat{S}_{ml} \;,     
\end{equation}
and in the case $S = 3/2$ the expansion ends with the
trivector term, and so on.

After this small digression let us come back to the derivation of the
Landau-Lifshitz equation. We have already noticed that it is necessary
to insert the Liouville equation (\ref{Liouville}) or the
Heisenberg equation (\ref{Heisenberg}) in Eq.~(\ref{STime}) to derive
a complete solvable equation for the expectation value $\langle
\hat{\mathbf{S}} \rangle$. The result is the following equation for
the components of $\hat{\mathbf{S}}$ ($n \in \{x,y,z\}$): 
\begin{eqnarray} \label{TraceSTime}
\mathrm{Tr}\left(\frac{\mathrm{d}}{\mathrm{d}t}\hat{\rho}\hat{S}_n\right)
= i\mathrm{Tr}\left([\hat{\rho},\hat{\mathrm{H}}]\hat{S}_n
\right) - \lambda\mathrm{Tr}
\left([\hat{\rho},[\hat{\rho},\hat{\mathrm{H}}]] \hat{S}_n\right) \;.
\end{eqnarray} 
Inserting $\hat{\rho}$ in (\ref{TraceSTime}) and calculating the
traces lead to an equation for $\langle \hat{\mathbf{S}} \rangle$
which is similar to the Landau-Lifshitz equation if the Heisenberg
Hamilton operator can be written as external field $\hat{\mathrm H} =
-\mathbf{B}_{\mathrm{eff}}\cdot\hat{\mathrm{S}}$. 

During the calculation we can ignore the first term of 
$\hat{\rho}$ because the identity matrix $\hat{\mathbf{1}}$ commutates
with any operator and therefore all the commutators with the identity
matrix become zero. Furthermore, we can see that only the vector term 
$\sum_m \langle \hat{S}_m \rangle \hat{S}_m$ participate to the
differential equation of $\langle \hat{\mathbf{S}} \rangle$. All the
higher order expansion terms (bivector, trivector, $\ldots$) are not
contributing. The reason for that is the conservation of the rank $k$ of
a tensor under rotation. Now, each term of the multivector expansion
has its own rank $k$. The expansion starts with the Identity matrix
with rank $k=0$ (scalar). The vector term has rank $k=1$, the
bivector rank $k=2$, the trivector rank $k=3$, and so on. The rank of
each additional term increases by a factor 1. 

Now, we assume that the trajectories of $\langle \hat{\mathbf{S}}
\rangle$ shall behave classical. This means that the length of the
spin is conserved, and the spin only fulfills precession and
relaxation. However, any precession as well as relaxation can be
described as a rotation of the coordinate system and we have already
mentioned that the rank of a tensor is conserved under such a
rotation. This also means that the motion of $\langle \hat{S}_n
\rangle$ is only described by the tensors $\hat{S}_n$ which are
responsible for $\langle \hat{S}_n \rangle$:
\begin{eqnarray} \label{SnExpValue}
\langle \hat{S}_n \rangle &=& \mathrm{Tr}(\hat{\rho}\hat{S}_n)
\nonumber \\
&=& \frac{1}{2S+1}\underbrace{\mathrm{Tr}(\hat{S}_n)}_{=0} +
\frac{1}{n_S}\sum\limits_m \langle \hat{S}_m \rangle
\underbrace{\mathrm{Tr}(\hat{S}_m\hat{S}_n)}_{=n_S\delta_{nm}}
\nonumber \\
&+& \frac{1}{n_{2S}}\sum\limits_{ml} \langle \hat{S}_{ml} \rangle
\underbrace{\mathrm{Tr}(\hat{S}_{ml}\hat{S}_n)}_{=0} + \underbrace{0 +
  \ldots + 0}_{\mathrm{Eq}.~(\ref{MegaTraceRule})} \;.
\end{eqnarray}        
The higher order expansion terms disappear due to
\begin{equation} \label{MegaTraceRule}
\mathrm{Tr}\left(\hat{S}_\alpha^k\hat{S}_\beta^{k'}\right) =
n_{S_k}\delta_{\alpha\beta}\delta_{kk'} \;.
\end{equation}
$k$ and $k'$ are the rank of the tensor and $\alpha$ and $\beta$ its
components: e.g. $\hat{S}_3^1 = \hat{S}_z$, or $\hat{S}_4^2 =
\hat{S}_{zx}$. 
  
Inserting the vector term $\sum_m \langle \hat{S}_m \rangle \hat{S}_m$
of $\hat{\rho}$ in Eq.~(\ref{TraceSTime}) we find after some algebra:
\begin{equation}
\frac{\mathrm{d}}{\mathrm{d}t}\langle \hat{S}_n \rangle = 
\left(\langle \hat{\mathbf{S}} \rangle \times
\mathbf{B}_{\mathrm{eff}} \right)_n
-\lambda\left(\langle \hat{\mathbf{S}} 
\rangle \times \left(\langle \hat{\mathbf{S}} 
\rangle \times \mathbf{B}_{\mathrm{eff}}\right)\right)_n \;.
\end{equation}
or written as complete vector:
\begin{equation}
\frac{\mathrm{d}}{\mathrm{d}t}\langle \hat{\mathbf{S}} \rangle = 
\langle \hat{\mathbf{S}} \rangle \times
\mathbf{B}_{\mathrm{eff}} 
-\lambda\langle \hat{\mathbf{S}} 
\rangle \times \left(\langle \hat{\mathbf{S}} 
\rangle \times \mathbf{B}_{\mathrm{eff}}\right) \;.
\end{equation}
This equation is similar to the classical Landau-Lifshitz (LL) equation
\cite{landauPZS32}:
\begin{equation} \label{EOM_LL}
\frac{\mathrm{d}}{\mathrm{d}t}\mathbf{S} = 
\mathbf{S} \times
\mathbf{B}_{\mathrm{eff}} -\lambda \mathbf{S} \times \left(\mathbf{S}
\times \mathbf{B}_{\mathrm{eff}}\right) \;.  
\end{equation}

A complete evidence of the previous statement 
and details of the calculation can be found in the
supplementary of this publication \cite{Suppl}.

Now, it is known that the LL equation
will lead to unphysical results in the limit of a large damping
$\lambda \gg 1$. The reason is that the precessional motion is not
influenced by the damping. The equation which holds in
the limit of a large damping is the Landau-Lifshitz-Gilbert (LLG)
equation. In the case of the LLG equation both terms the precessional
term as well as the relaxation term are influenced by the damping. The
easiest way to obtain the LLG equation from the LL equation is to make
the following transformation of the time $t$: $t \rightarrow
t/(1+\lambda^2)$. The same transformation in the case of Schr\"odinger
equation (\ref{TDSE_LL}) leads to:
\begin{eqnarray} \label{TDSE_LLG}
i(1+\lambda^2)\frac{\mathrm{d}}{\mathrm{d}t}|\psi(t)\rangle =
(\hat{\mathrm{H}} - i \lambda [\hat{\mathrm{H}} - 
  \langle \hat{\mathrm{H}} \rangle ])|\psi(t)\rangle \;.
\end{eqnarray}
This equation can be seen as the quantum mechanical analog of the LLG
equation: 
\begin{equation} \label{clLLG}
\frac{\mathrm{d}}{\mathrm{d}t}\langle \hat{\mathbf{S}} \rangle = 
\frac{1}{1+\lambda^2}\langle \hat{\mathbf{S}} \rangle \times
\mathbf{B}_{\mathrm{eff}} 
-\frac{\lambda}{1+\lambda^2}\langle \hat{\mathbf{S}} 
\rangle \times \left(\langle \hat{\mathbf{S}} 
\rangle \times \mathbf{B}_{\mathrm{eff}}\right) \;.    
\end{equation}
The classical LLG equation itself appears by replacing the spin
expectation value $\langle \hat{\mathbf{S}} \rangle$ by the classical spin
$\mathbf{S}$. 

\section{Single spin vs. multi-spin system} \label{s:oneormore}

This could be the end of the story. However, there is
one point which has not been discussed so far. In the classical
description we assume a constant length of the spin: $|\mathbf{S}| =
1$. The LLG equation at zero temperature just describes a precessional
motion resp. relaxation due to an effective field
$\mathbf{B}_{\mathrm{eff}}$. In the quantum mechanical description we
deal with the expectation values $\langle \hat{\mathbf{S}} \rangle =
\langle \psi | \hat{\mathbf{S}} | \psi \rangle$ which strongly depend
on the wave functions $|\psi \rangle$. The absolute value $|\langle
\hat{\mathbf{S}} \rangle|$ of these expectation values have not
necessarily to be constant: $|\langle \hat{\mathbf{S}} \rangle| \leq
\hbar S$. Furthermore, due to the time dependence of the wave function
$|\psi \rangle = |\psi(t) \rangle$, $|\langle \hat{\mathbf{S}}
\rangle|$ is also time dependent. This means that in general we cannot
expect an agreement of the classical with the quantum mechanical spin
dynamics. On the other hand, we find in textbooks the statement that
we can expect a classical behavior, at least for a single spin in an
external field and without damping \cite{guoBOOK08}. This textbook
statement of a semiclassical description of the spin dynamics is not totally
wrong. In the case of single spins and in the case of linear
excitations of spin system in the ferromagnetic ground state (spin
wave excitation) the classical and the quantum mechanical description
can lead to the same trajectories if the Hamiltonian is linear in
$\hat{\mathbf{S}}_n$ ($n$ stands for the lattice site and not for a
spin component. For details see \cite{wieserPRB11}).    

To specify the description let us first assume a single spin which is
in a pure state $|\psi(t) \rangle$. In this case the density operator
is given by Eq.~(\ref{PureDen}) and
we find $\mathrm{Tr}(\hat{\rho}^2) = \mathrm{Tr}(\hat{\rho}) = 1$. For
comparison, in the case of a mixed state:
\begin{equation}
\hat{\rho} = \sum_i p_i |\psi_i
\rangle \langle \psi_i | \;
\end{equation} 
we find $\hat{\rho}^2 \neq \hat{\rho}$, therefore
$\mathrm{Tr}(\hat{\rho}^2) < \mathrm{Tr}(\hat{\rho}) = 1$. Then, we
can write $\mathrm{Tr}(\hat{\rho}^2)$ in the case of a single spin
with $S = 1/2$:
\begin{eqnarray}
\mathrm{Tr}(\hat{\rho}^2) = \frac{1+ |\langle
  \hat{\boldsymbol{\sigma}} \rangle|^2}{2} \leq 1 \;. 
\end{eqnarray}
A similar formula can be given for higher spin quantum numbers ($S >
1/2$) due to Eq.~(\ref{MegaTraceRule}). However, in these cases it is
impossible to give a geometric plot as for $S = 1/2$ (Bloch sphere). 

$\mathrm{Tr}(\hat{\rho}^2) = 1$ appears in the case of a pure state
which corresponds to $|\langle \hat{\boldsymbol{\sigma}} \rangle| = 1$
resp. $|\langle \hat{\mathbf{S}} \rangle| = \hbar/2$. In the case of a
spin ensemble in a mixed state we have $\mathrm{Tr}(\hat{\rho}^2) < 1$ 
corresponding to $|\langle \hat{\mathbf{S}} \rangle| < \hbar/2$. This
means, in the case of a single spin in a pure state the norm of the
expectation value $|\langle \hat{\mathbf{S}} \rangle|$ is constant and
the dynamics can become classical [see Eq.~(\ref{clLLG})].

In the case of a multi-spin system the wave function is described by a
product of the eigenstates of the single spins:
\begin{equation} \label{ProdState}
|\psi \rangle = |S_1 m_{1}\rangle \otimes |S_2 m_{2}\rangle \otimes
\ldots \otimes  |S_N m_{N}\rangle \;.
\end{equation}
Such a state is called product state. For example the
ferromagnet state is described by a product state: 
$|\mathrm{FM} \rangle = |\uparrow \uparrow \ldots \uparrow \rangle  =
|\uparrow \rangle \otimes |\uparrow \rangle \otimes \ldots \otimes |\uparrow
\rangle$. 

However, product states are normally not the general states of a
magnetic system. The normal state in a quantum mechanical system with
more then one spin is a superpositions of the product states
(\ref{ProdState}):  
\begin{equation} \label{waveMulti}
|\psi \rangle = \sum\limits_{m_1\ldots m_N} c_{m_1m_2\ldots m_N}
\left(|S_1 m_{1}\rangle \otimes |S_2 m_{2}\rangle \otimes \ldots
\otimes |S_N m_{N}\rangle \right)\;.
\end{equation} 
The product states in this case cannot be separated which that we
cannot write $c_{m_1m_2\ldots m_N} =
c_{m_1}c_{m_2}\ldots c_{m_N}$. This means that we have a
correlation between the spins which do not exist in
classical physics. This phenomenon is called entanglement
\cite{garaninAdvChemPhys11,krammerPHDThesis} and leads to the fact
that in the case of two entangled spins a measurement on the first
spin also affects the second spin \cite{einsteinPR35}. The same is
true for more then two spins. In this case the measurement on one spin
affects all spins. Only in the case of a product state the subsystems
are not correlated and can be separated, which means that the
measurement of a single spin in this case only affect this spin and do
not change the wave functions of the other spins as in the case of
entanglement.      

Now, the expectation value $\langle \hat{\mathbf{S}}_n \rangle$, where $n$
is the lattice site, can be calculated similar to a single spin by:
\begin{equation} \label{EC1}
\langle \hat{\mathbf{S}}_n \rangle = \langle \psi |\hat{\mathbf{S}}_n |
\psi \rangle \;,
\end{equation} 
where $| \psi \rangle$ is given by (\ref{ProdState}) or
(\ref{waveMulti}), or by: 
\begin{equation} \label{EC2}
\langle \hat{\mathbf{S}}_n \rangle =
\mathrm{Tr}\left(\hat{\rho}\hat{\mathbf{S}}_n\right) \;.
\end{equation}
In the following we assume a pure state. Therefore, the density operator of
our multi-spin system $\hat{\rho}$ is defined by $\hat{\rho} =
|\psi \rangle \langle \psi |$.

Alternativ we can calculate the expectation value $\langle
\hat{\mathbf{S}}_n \rangle$ with aid of the reduced density operator
$\hat{\rho}_{m_n}$ \cite{garaninAdvChemPhys11}:
\begin{equation} \label{EC3}
\langle \hat{\mathbf{S}}_n \rangle =
\mathrm{Tr}_n\left(\mathrm{Tr}_{j\neq
  n}\left(\hat{\rho}\hat{\mathbf{S}}_n\right)\right) =
\mathrm{Tr}_n\left(\hat{\rho}_{m_n}\hat{\mathbf{S}}_n\right)\;,  
\end{equation} 
where $\mathrm{Tr}_n$ is the partial trace over the sub Hilbert space
$\mathcal{H}_n$ and $\mathrm{Tr}_{j \neq n}$ the partial trace over
all the other sub Hilbert spaces: $\mathcal{H} = \mathcal{H}_1 \otimes
\mathcal{H}_2 \otimes \ldots \otimes \mathcal{H}_N$. The advantage of
using the reduced density operator $\hat{\rho}_{m_n}$ is that we have
to deal with smaller matricies, which reduces the numerical effort.

It is easy to verify that the reduced density operator $\hat{\rho}_{m_n}$
is given by:
\begin{equation} \label{RedDen}
\hat{\rho}_{m_n} = \sum\limits_{m_n} p_{m_n} |S_n m_{n}\rangle \langle S_n
m_{n} | \;.
\end{equation}
Eq.~(\ref{RedDen}) clearly shows that the reduced density operator
corresponds to a mixed state, the $p_{m_n}$ are the probabilities to find
the system in the state $|S_n m_n \rangle$, even if the complete
system is in a pure state $\hat{\rho} = |\psi \rangle \langle \psi |$.   

In the case of the product state (\ref{ProdState}) the reduced density
operator is given by $\hat{\rho}_{m_n} = |S_n m_{n}\rangle \langle S_n
m_{n} |$. Here, the reduced density operator $\hat{\rho}_{m_n}$ as
well as the density operator of the complete system $\hat{\rho} =
|\psi \rangle \langle \psi |$ are pure.  

The consequences are as long as the system is described by a 
product state we find: $|\langle \hat{\mathbf{S}}_n \rangle| = \hbar
S$. However, as mentioned before the normal states are superposition
states which are entangled and not product states. In these cases we
find $|\langle \hat{\mathbf{S}}_n \rangle| < \hbar S$ (see discussion
before about the density operators of pure and mixed states). The
explicit value of $|\langle \hat{\mathbf{S}}_n \rangle|$ corresponds
to the strength of the entanglement: no entanglement means $|\langle
\hat{\mathbf{S}}_n \rangle| = \hbar S$ which is the maximal value and
with entanglement $|\langle \hat{\mathbf{S}}_n \rangle|$ decreases and
can become zero. Due to the fact that the entanglement depends on
$|\psi \rangle$ and $|\psi \rangle$ can change with the time $t$, $|\langle
\hat{\mathbf{S}}_n \rangle|$ can also change with the time.     

Another way to quantify the entanglement is the von Neumann entropy:
\begin{equation} \label{vNEn}
S(\hat{\rho}_{m_n}) =
-\mathrm{Tr}\left(\hat{\rho}_{m_n}\mathrm{log}_2\hat{\rho}_{m_n}
\right) \;,  
\end{equation}
where $\hat{\rho}_{m_n}$ is the reduced density matrix. 

The von Neumann entropy is the extension of the classical Gibbs
entropy concept to the quantum mechanics and can be seen as the
quantum mechanical analog to the classical Shannon entropy 
\cite{shannonBSTJ48} of the information technology.

In principle the von Neumann entropy is defined for any density
matrix $\tilde{\rho}$. In general, the von Neumann entropy proves if
the system is in a pure or in a mixed state. Therefore, it makes sense
to use the reduced density matrix $\hat{\rho}_{m_n}$ because in these
cases a pure state corresponds to a product state, and therefore to no
entanglement. In these cases we find $S(\hat{\rho}_{m_n}) = 0$. A
mixed state corresponds to an entangled superposition state. In these
cases we find $0 < S(\hat{\rho}_{m_n}) \leq 1$. The maximum value
$S(\hat{\rho}_{m_n}) = 1$ corresponds to $|\langle \hat{\mathbf{S}}_n
\rangle| = 0$. Furthermore, as before $|\langle \hat{\mathbf{S}}_n \rangle|$,
$S(\hat{\rho}_{m_n})$ is time dependent.    
  
\section{NUMERICAL PROOF} \label{s:simulations}
In the previous section we have seen that only in the cases of a
single spin or a spin ensemble which behaves like a single spin and in
the cases of a linear excitation around the ferromagnetic ground state
we can find a quantum mechanical spin dynamics similar to the classical
spin dynamics. In these cases we have no entanglement and the absolute
value $|\langle \hat{\mathbf{S}} \rangle |$ stays constant. In the
cases of entanglement this is not the case. To proof this statement
computer simulations have been performed. In the previous 
publications only linear excitations \cite{wieserPRB11} or single
spins \cite{wieserPRL13} have been investigated. In the 
following, we will discuss the magnetization reversal of a single spin
as well as a linear excitation and a magnetization reversal of a trimer. 

The Heisenberg Hamiltonian of our
system is given by: 
\begin{eqnarray} \label{Ham}
\hat{\mathrm{H}} = -J\sum\limits_n
\hat{\mathbf{S}}_n\hat{\mathbf{S}}_{n+1} 
-\mu_S \sum\limits_n B_z \hat{S}_n^z + -\mu_S B_x(t) \hat{S}_1^x \;.
\end{eqnarray}     
The first term describes the nearest neighbor exchange interaction with
ferromagnetic $(J > 0)$ or antiferromagnetic $(J < 0)$ coupling. 

The second sum and third term describe couplings to external magnetic
fields. In this case a static field in $z$-direction and a field pulse
in $x$-direction acting on the first spin only: 
\begin{equation}
B_x(t) = B_0^xe^{-\frac{1}{2}\left(\frac{t-t_0}{T_W}\right)^2}\;.
\end{equation}
This field pulse will be used to bring the system out of equilibrium
and to start the reversal process or just for linear excitations.

For the moment we restrict ourself to a single spin [no exchange term
in (\ref{Ham})]. The spin is initially oriented in $+z$ direction:
$|\psi(t=0)\rangle = |\uparrow \rangle$. The static 
external field in $-z$-direction. The field pulse is needed to break
the symmetry and to initialize the magnetization reversal.   

\begin{figure}[h]
  \includegraphics*[width=6.cm,bb = 70 380 545 725]{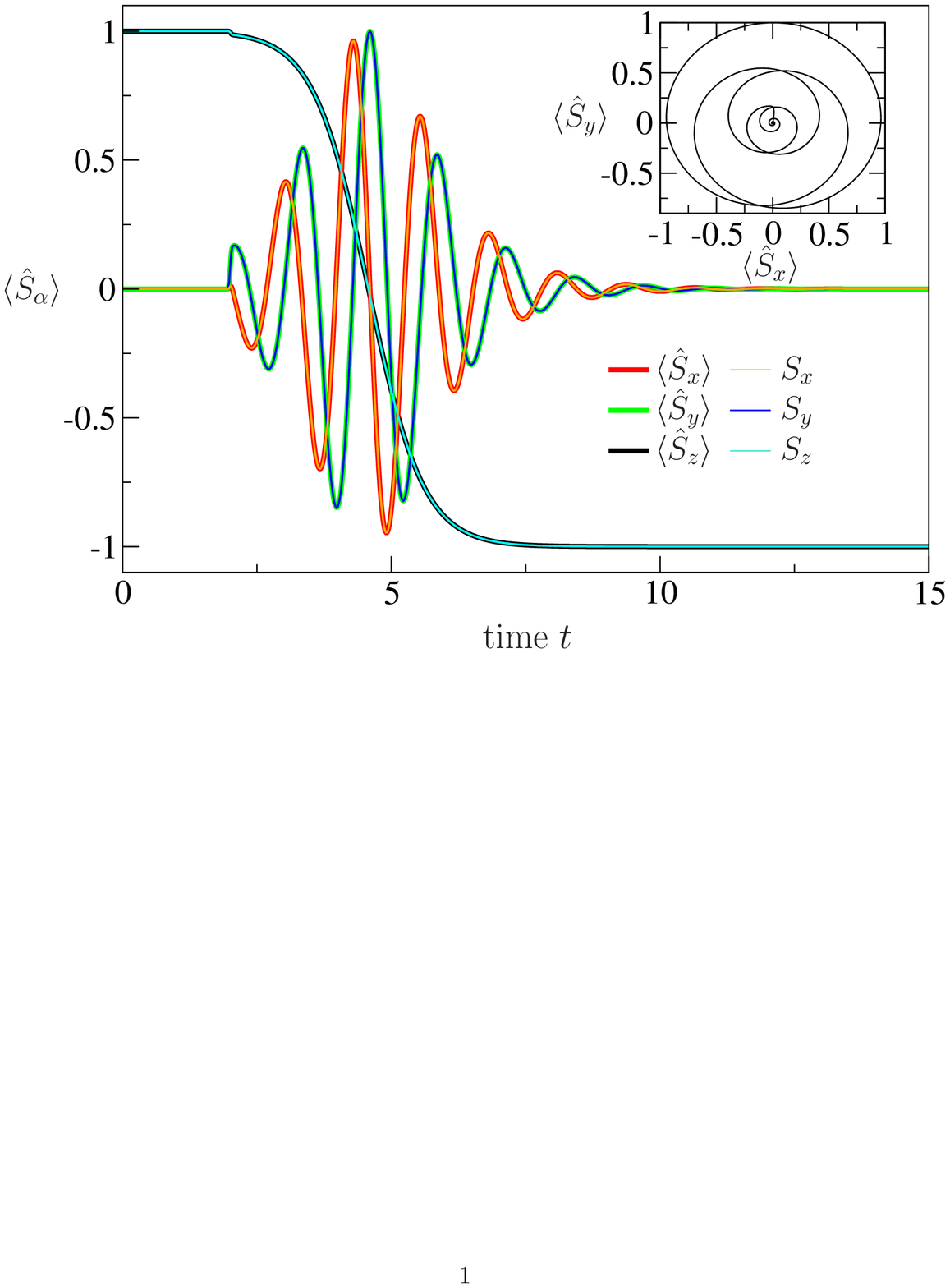}
  \caption {Magnetization reversal of a single spin. For comparison
    the quantum mechanical expectation values $\langle \hat{S}_n
    \rangle$ $n \in \{x,y,z\}$ and the classical spin components have
    been plotted as function of time. The inset shows the precessional
    motion during the reversal. 
    ($S = 1$, $J = 0$, $\mu_SB_z = -5.1$, $\mu_SB_0^x = 3.27$, $t_0 =
    2$, $T_W = 0.02$, $\lambda = 0.2$)}      
  \label{f:pic1}
\end{figure}
Fig.~\ref{f:pic1} shows the trajectories of $\langle \hat{\mathbf{S}}
\rangle$ calculated with the Schr\"odinger equation (\ref{TDSE_LLG})
and for comparison the trajectory of the classical spin $\mathbf{S}$
calculated with the classical Landau-Lifshitz-Gilbert equation. As
expected, Fig.~\ref{f:pic1} shows a perfect agreement of both, the
classical as well as quantum mechanical trajectories, which also means
that $|\langle \hat{\mathbf{S}} \rangle|$ is conserved. For a detailed
discussion please see Sec.~\ref{s:LLG} and \ref{s:oneormore}.   

Let us come to the trimer. All spins are aligned in a chain and we
assume that there is no coupling between the first and the last
spin. The Hamiltonian (\ref{Ham}) is linear in 
$\hat{\mathbf{S}}_n$ and therefore we can expect a classical behavior
if we are close to the ferromagnetic ground state. 

\begin{figure}[h]
\includegraphics*[width=6.cm,bb = 70 380 545 725]{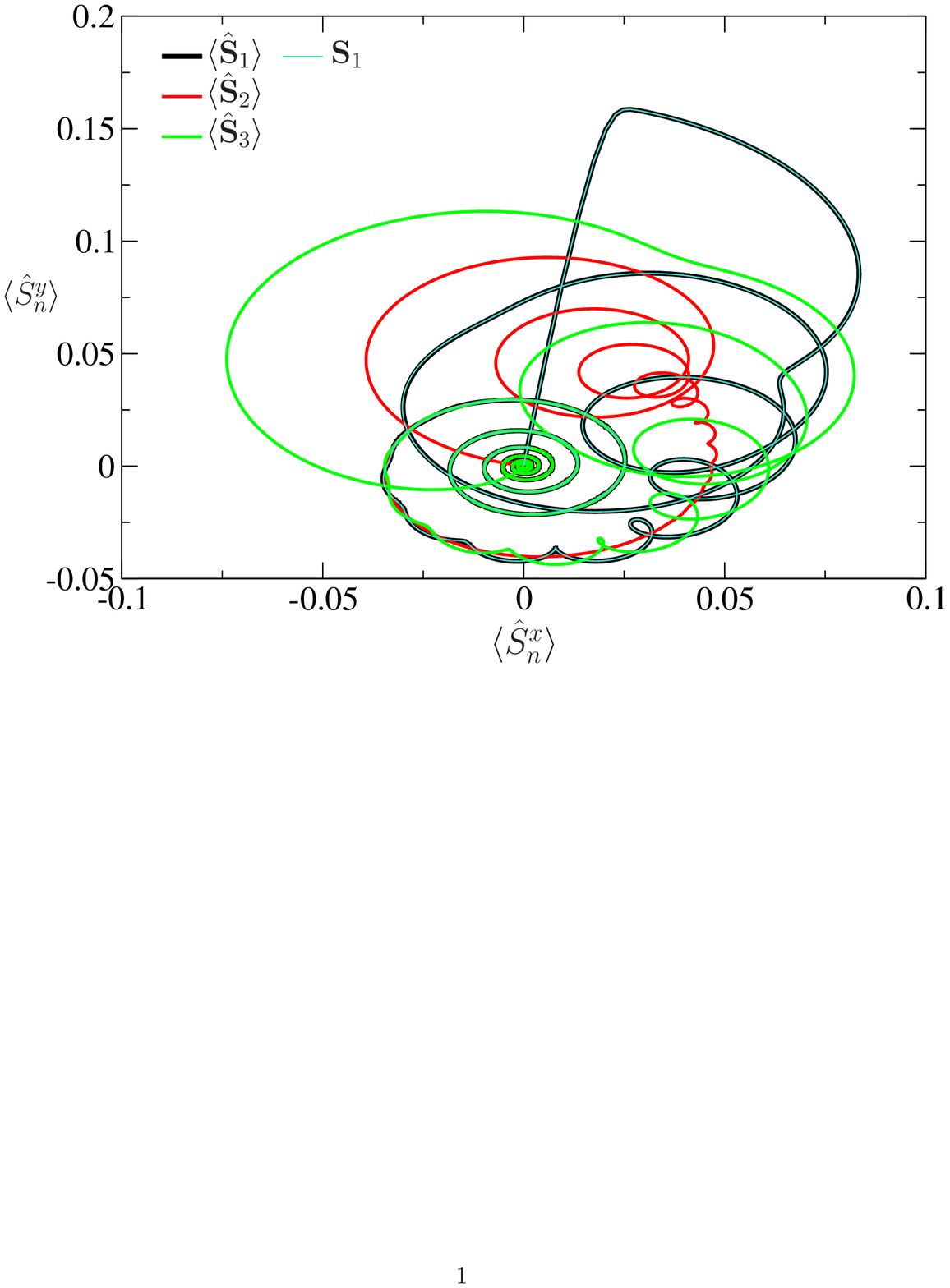}  
  \caption {Linear excitation and additional relaxation of three
    ferromagnetic coupled spins: only the first spin has been excited
    by the field pulse. A perfect agreement between quantum mechanical and
    classical trajectory can be seen for the first spin. The same
    is true also for the other spins. 
    ($S = 1$, $J = 1$, $\mu_SB_z = 0.1$, $\mu_SB_0^x = 3.27$, $t_0 =
    10$, $T_W = 0.02$, $\lambda = 0.1$)}       
  \label{f:pic2}
\end{figure}

Fig.~\ref{f:pic2} shows the trajectories of three ferromagnetic
coupled spins, excited by a tiny field pulse (linear excitation), with
$S = 1$. The field pulse acts on the first spin only. In the initial 
configuration is: all spins are oriented in $+z$-direction, the direction
of the external field $B_z$. 

To show the agreement of the classical spins $S_n$ with the
expectation values $\langle \hat{\mathbf{S}}_n 
\rangle$ we have compared the curves of the first spin: it can be seen
that both curves lie on top of each other which means that the
expectation values behave classical. The comparison between classical
and quantum mechanical trajectory of the other two spins show the same
behavior (not shown).  

As mentioned before such an agreement can be expected only for linear
excitations of the ferromagnetic ground state, in this case
$|\mathrm{FM}\rangle = |\uparrow\uparrow\uparrow\rangle$. This is the
case within the simulation. This can be seen by the small amplitude
of the $\langle \hat{S}_n^x \rangle$ and $\langle \hat{S}_n^y \rangle$
component. Furthermore, the simulation shows that only the
basis state $|S m_1\rangle \otimes |S m_2\rangle \otimes |S m_3\rangle
= |m_1m_2m_3 \rangle = |\uparrow\uparrow\uparrow\rangle$ gives a
relevant impact: $|\langle \uparrow\uparrow\uparrow | \psi(t) \rangle|^2 
\approx 1$. All other basis states $|m_1m_2m_3 \rangle$
are not occupied or just marginal with $|\langle m_1m_2m_3 | 
\psi(t) \rangle|^2 < 0.01$. Furthermore, additional simulations show
that with increasing $S$ the deviation from $|\psi(t) \rangle =
|\mathrm{FM}\rangle$, up to which we can see a classical behavior,
increases . This becomes clear because with $S \rightarrow \infty$ we 
find the classical limit.   
\begin{figure}[h]
\includegraphics*[width=4.cm,bb = 40 380 545 725]{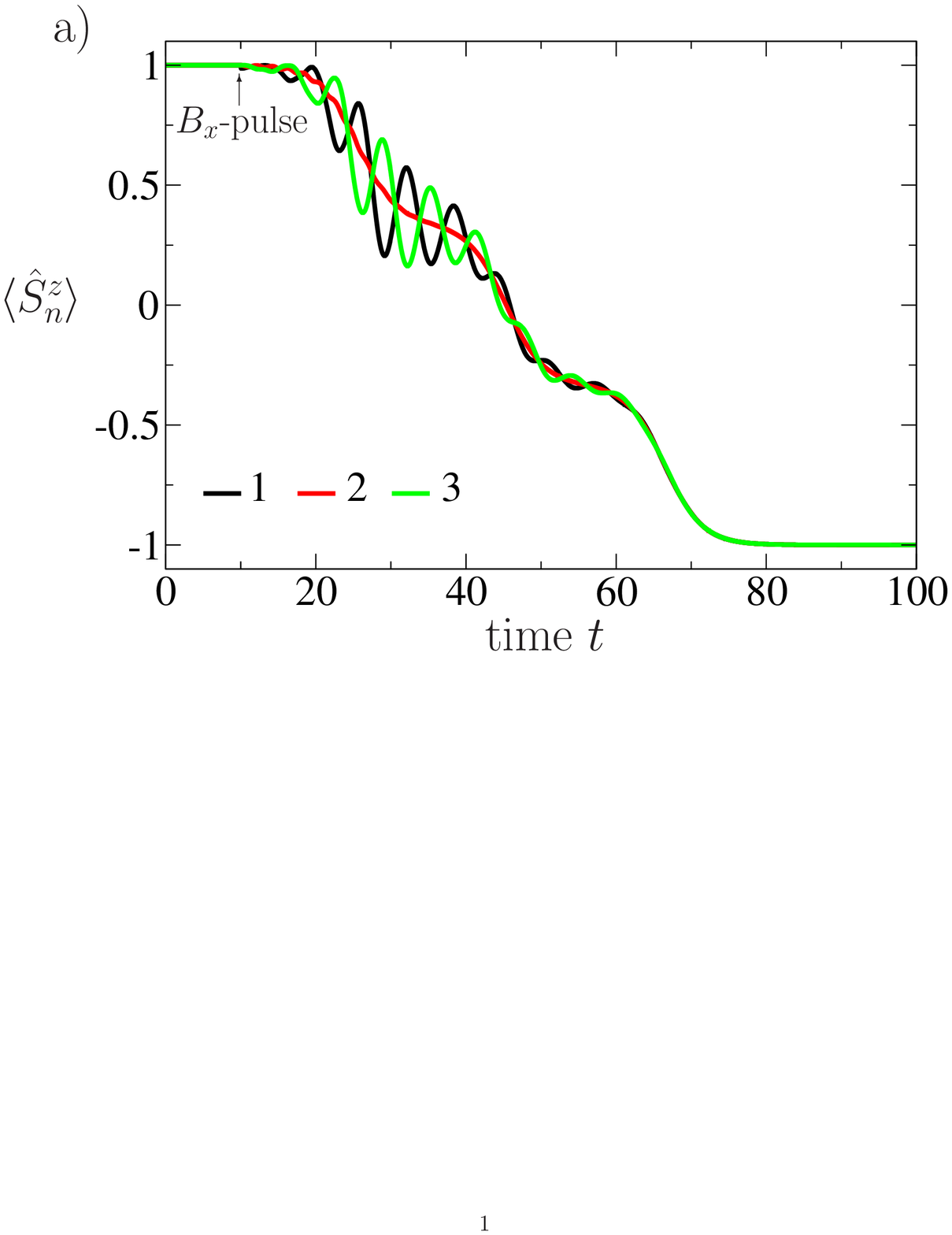}
\includegraphics*[width=4.cm,bb = 40 380 545 725]{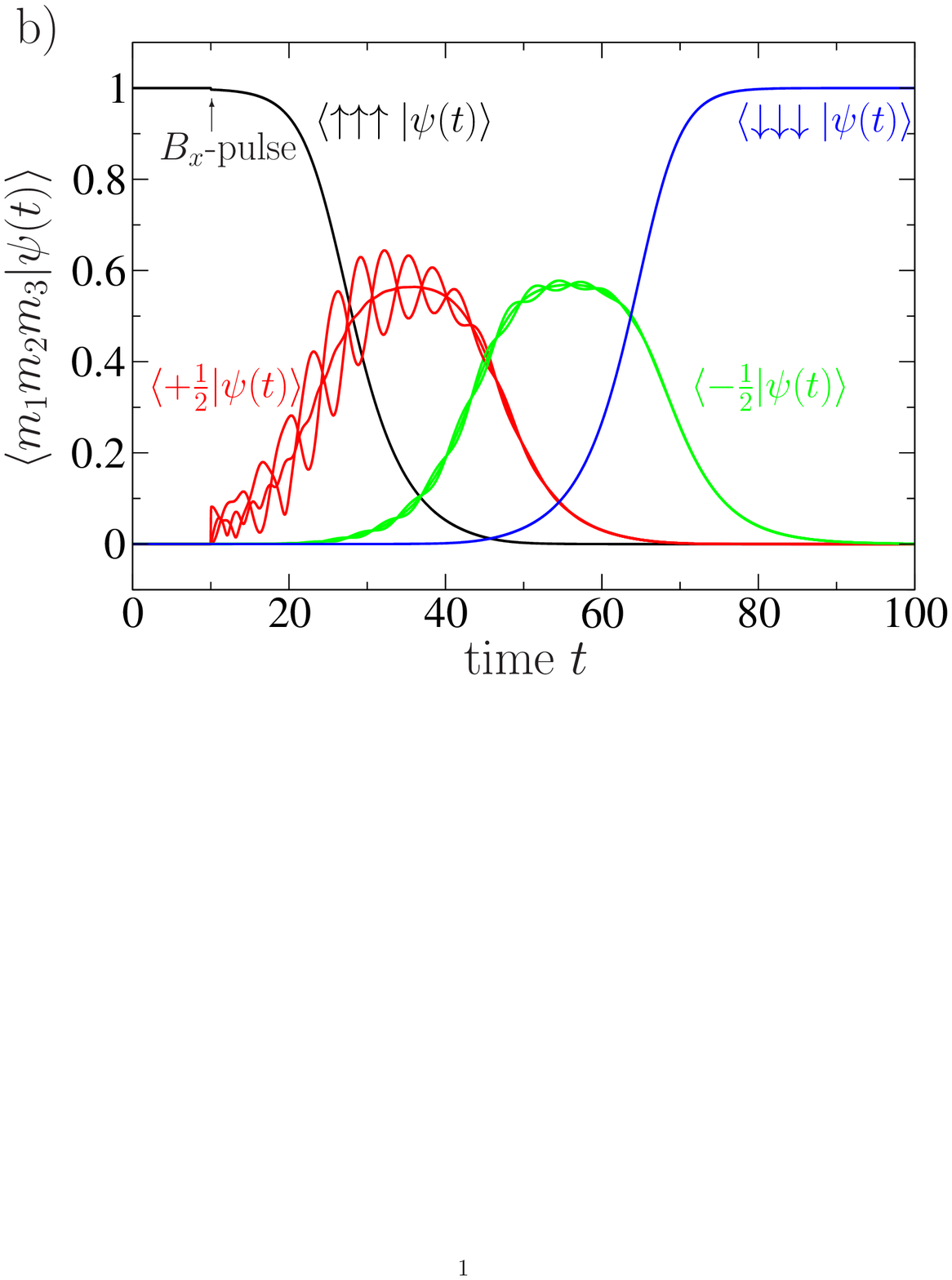} 
\includegraphics*[width=4.cm,bb = 40 380 545 725]{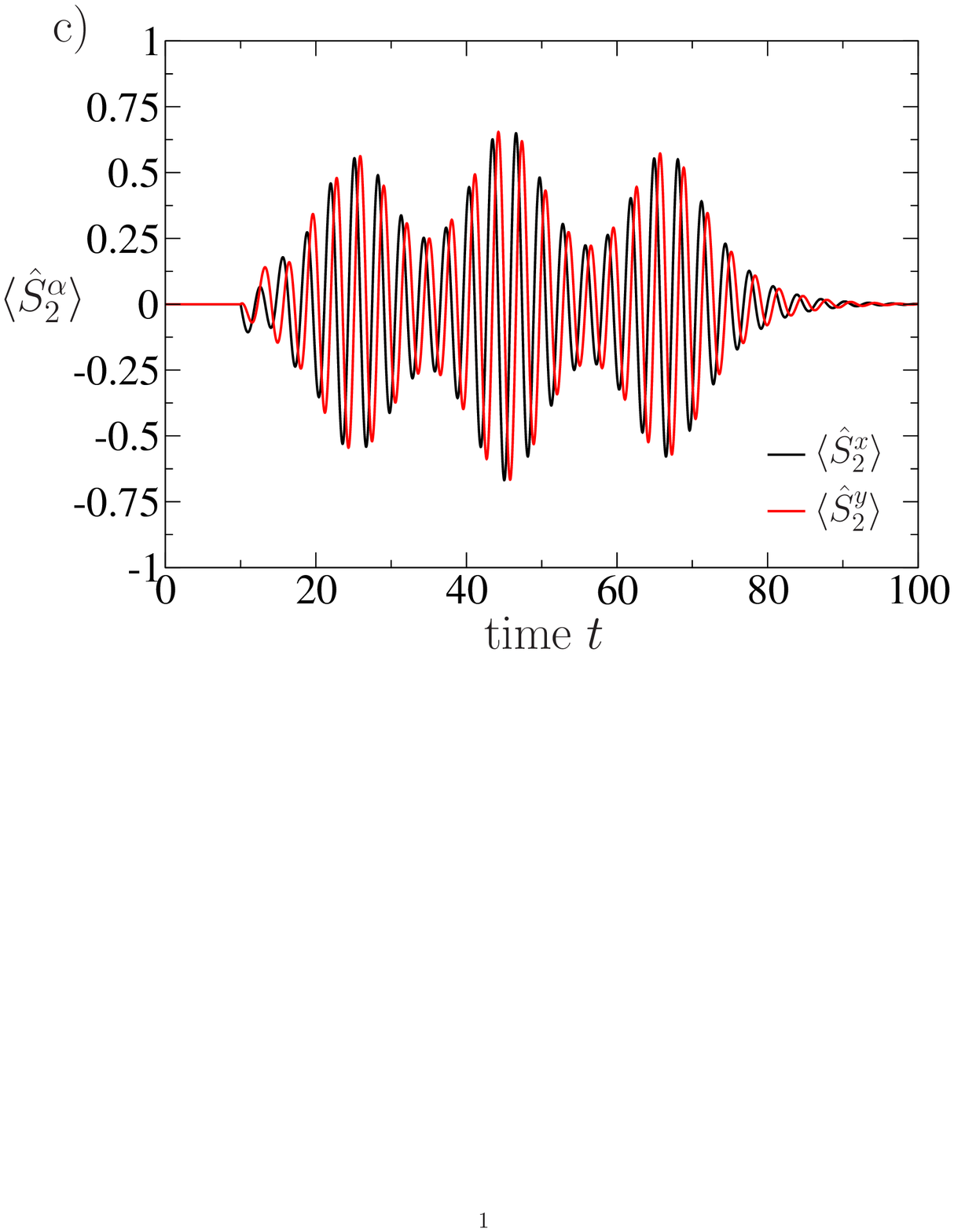}
\includegraphics*[width=4.cm,bb = 40 380 545 725]{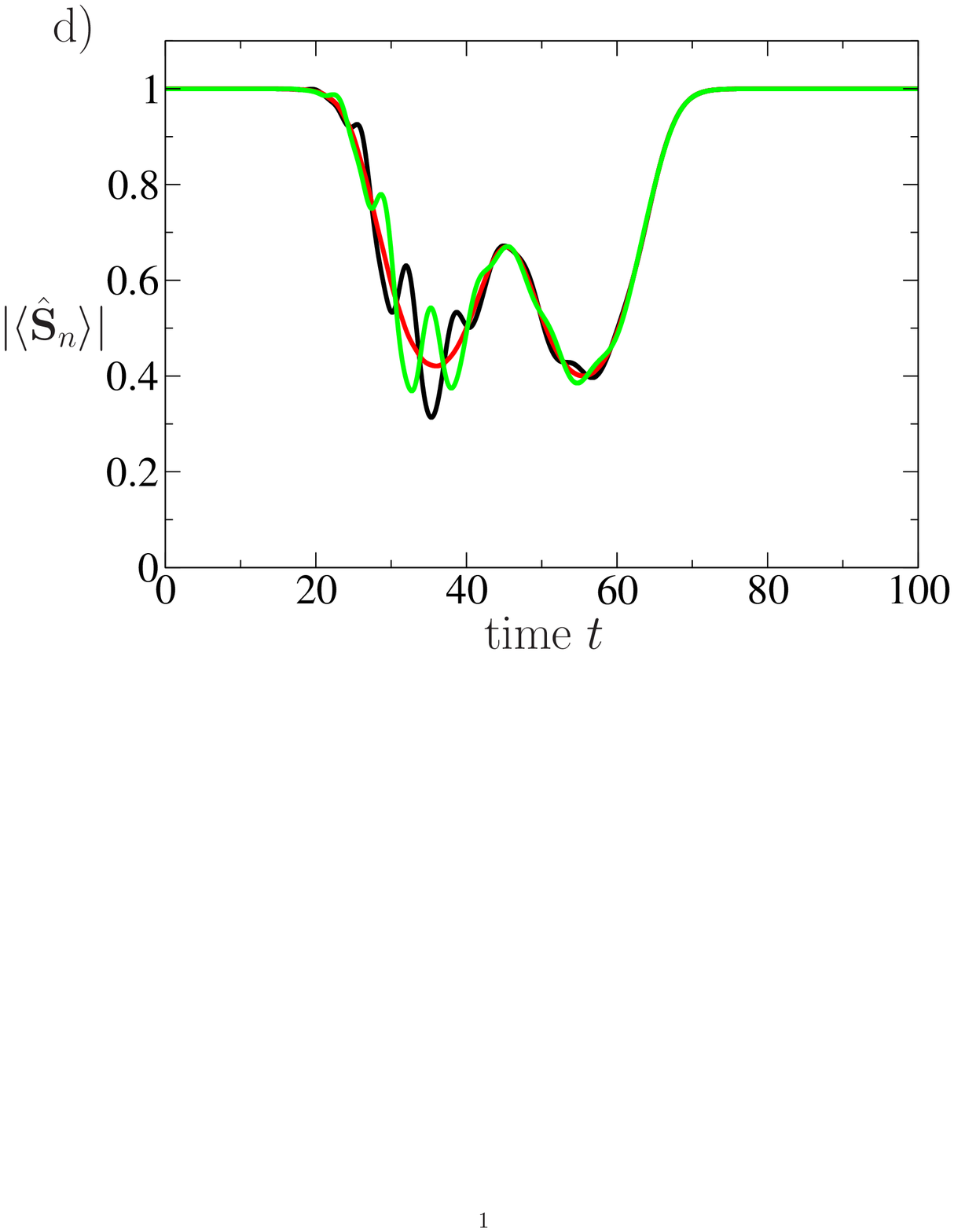}
  \caption {Magnetization reversal for three ferromagnetic couple
    spins. a) $z$-component of the spin as function of time $t$, b)
    projection of the wave function $|\psi(t) \rangle$ to the basis
    states $|m_1m_2m_3\rangle$, c) precession motion of the second
    spin during the reversal process, d) spin 
    length as function of time $t$. 
    ($S = 1/2$, $J = 4$, $\mu_SB_z = -2$, $\mu_SB_0^x = 3.27$, $t_0 =
    10$, $T_W = 0.02$, $\lambda = 0.1$)}       
  \label{f:pic3}
\end{figure}
\begin{figure}[h]
  \includegraphics*[width=6.cm,bb = 70 380 545 725]{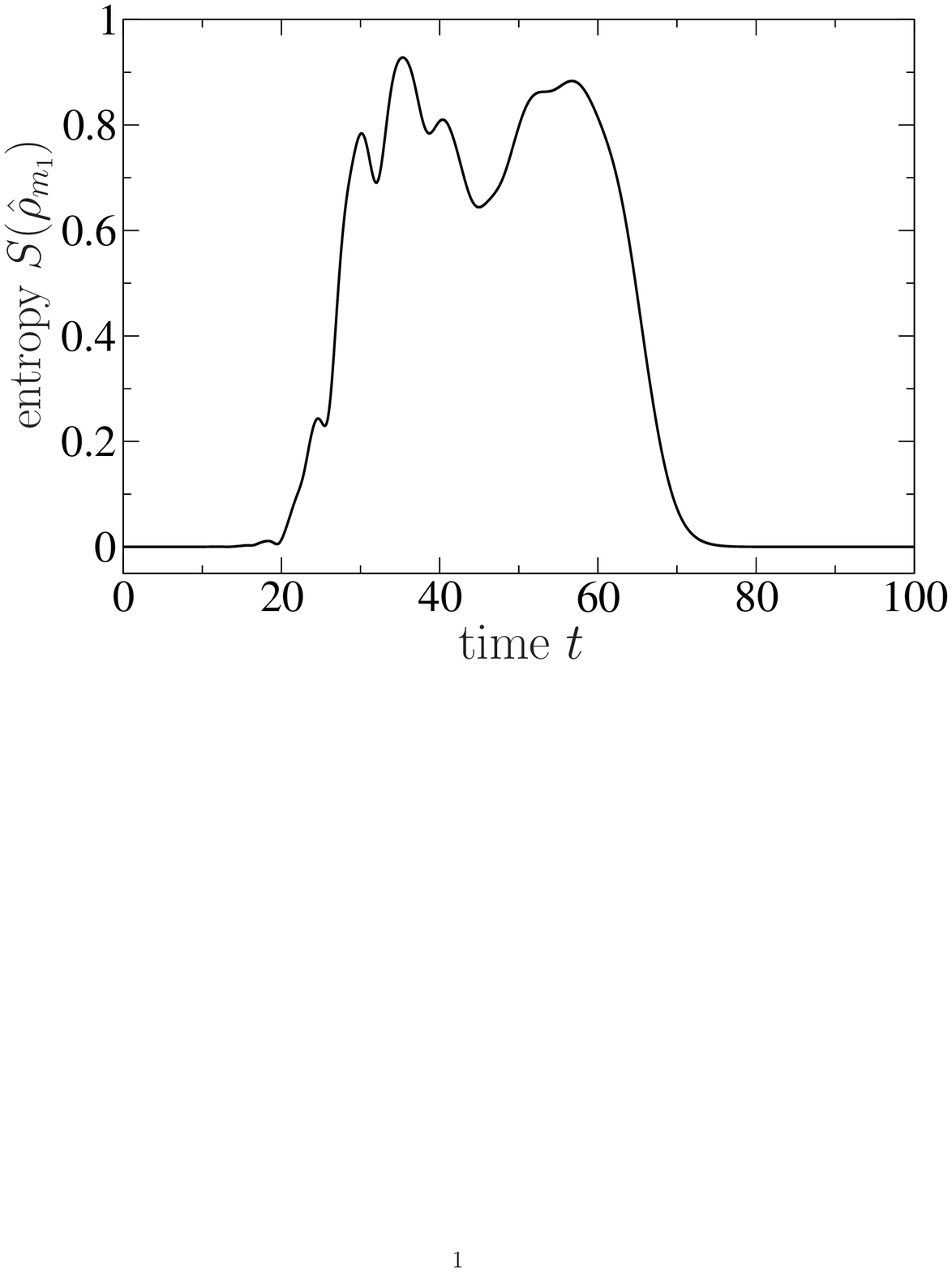}
  \caption {Von Neumann entropy $S(\hat{\rho}_{m_1})$ vs. time $t$,
    corresponding to the magnetization reversal of the three spin
    system presented in Fig.~\ref{f:pic3}.}      
  \label{f:pic4}
\end{figure}

The situation changes if we leave the ferromagnetic ground
state. Fig.~\ref{f:pic3} shows the magnetization reversal of three
coupled spins with $S = 1/2$. The initial configuration is the same as
before all three spins in $+z$-direction, however the external field
is now $-z$-direction. Again, the field pulse excites the first spin
only. The result is a stepwise magnetization reversal [see
Fig.~\ref{f:pic3}a)]. Each step corresponds to the occupation of one of
the basis states $|m_1m_2m_3 \rangle$ corresponding to 
$|+3/2\rangle = |\uparrow \uparrow \uparrow \rangle$ (basis state with
three spins up), $|+1/2\rangle$ (all basis states with two spins up),
$|-1/2\rangle$ (all basis states with two spins down), and
$|-3/2\rangle = |\downarrow \downarrow \downarrow \rangle$ (basis
state with three spins down) [see Fig.~\ref{f:pic3}b)]. The
oscillation of $\langle \hat{S}_1^z \rangle$ and $\langle \hat{S}_3^z
\rangle$ around $\langle \hat{S}_2^z \rangle$ is a direct result of
the field pulse acting on the first spin only and the exposed position
of the second spin as the middle of this three spin cluster. 

During the reversal process $\langle \hat{S}_n^x \rangle$ and $\langle
\hat{S}_n^y \rangle$ [Fig.~\ref{f:pic3}c)] show a precessional motion
of the spins similar to the magnetization reversal of the single spin
before [see Fig.~\ref{f:pic1}]. 

The most important result is shown in Fig.~\ref{f:pic3}d):
the not conserved absolut value $|\langle \hat{\mathbf{S}}_n 
\rangle|$. This means, we cannot expect an agreement
with the classical trajectories where we assume a constant spin length
$|\mathbf{S}_n| = 1$. As mentioned before $|\langle  
\hat{\mathbf{S}}_n \rangle| < \hbar S$ means that the system is
entangled, which can be seen also by Fig.~\ref{f:pic3}b). The three
spins are in a product state (FM) only at the beginning and at the end
of the reversal process. In between, we see superpositions
of the basis states $|m_1m_2m_3 \rangle = |S m_1 \rangle \otimes |S m_2 \rangle
\otimes |S m_3 \rangle$ which means entanglement.    

To strengthen this message the von Neumann entropy 
\begin{equation} \label{vonNeumannEntropy}
S(\hat{\rho}_{m_1}) =
-\mathrm{Tr}\left(\hat{\rho}_{m_1}\mathrm{log}_2\hat{\rho}_{m_1}
\right) \;,
\end{equation}
for the reduced density operator $\hat{\rho}_{m_1}$, corresponding with the
Hilbert space of the first spin of our three spin system, has been
calculated. For the details of the calculation please see the
supplementary material \cite{Suppl}. 

Fig.~\ref{f:pic4} shows the von Neumann entropy $S(\hat{\rho}_{m_1})$ as
function of time $t$. It can be seen that ferromagnetic states
at the beginning and end: $|\uparrow\uparrow\uparrow\rangle$ and
$|\downarrow\downarrow\downarrow\rangle$ show no
entanglement. Furthermore, we see that the highest entanglement
[max. $S(\hat{\rho}_{m_1})$] appears when the $|\langle
\hat{\mathbf{S}}_n \rangle |$, $n \in \{1,2,3\}$, have 
their smallest values.

\section{TEMPERATURE} \label{s:Temperature}

The previous sections describe the quantum spin dynamics with energy
dissipation at zero temperature. Within this section we discuss two
possibilities to include temperature effects. The easiest way is to
add to the Hamilton operator $\hat{\mathrm{H}}$ [Eq.~(\ref{Ham})] an
additional stochastic field term:
\begin{equation}
\hat{\mathrm{H}}_\xi = -\sum\limits_n \boldsymbol{\xi}_n \hat{\mathbf{S}}_n\;.
\end{equation}   
$\boldsymbol{\xi}_n = (\xi_n^x,\xi_n^y,\xi_n^z)$ is a stochastic
field characterized by  
$\langle \xi_n^\alpha(t) \rangle = 0$, and
$\langle \xi_n^\alpha(t)\xi_m^\beta(t') \rangle = D
\delta_{nm}\delta_{\alpha\beta} \delta(t-t')$, with $\alpha,\beta \in
\{x,y,z\}$, and lattice sites $n$,$m$. Depending on the prefactor $D$
the stochastic field can be used to describe temperature or quantum
fluctuations \cite{pokrovskyPRB03}.  

The advantage of such an stochastic field is that this term is already
a field term. This means that this term behaves classical. In other
words: the Heisenberg Hamiltonian $\hat{\mathrm{H}} =
-(\mathbf{B}_{\mathrm{eff}}+ \boldsymbol{\xi})
\hat{\mathbf{S}}$ inserted in the Schr\"odinger equation
(\ref{TDSE_LLG}) immediately leads to the following Langevin equation
which can be seen as the quantum mechanical analog of the stochastic
Landau-Lifshitz-Gilbert equation \cite{garciaPRB98,nowakARCP01}: 
\begin{eqnarray}
\frac{\mathrm{d}}{\mathrm{d}t}\langle \hat{\mathbf{S}} \rangle &=& 
\frac{1}{1+\lambda^2}\langle \hat{\mathbf{S}} \rangle \times
\mathbf{B}_{\mathrm{eff}} 
-\frac{\lambda}{1+\lambda^2}\langle \hat{\mathbf{S}} 
\rangle \times \left(\langle \hat{\mathbf{S}} 
\rangle \times \mathbf{B}_{\mathrm{eff}}\right) \nonumber \\
&+& \frac{1}{1+\lambda^2}\langle \hat{\mathbf{S}} \rangle \times
\boldsymbol{\xi} 
-\frac{\lambda}{1+\lambda^2}\langle \hat{\mathbf{S}} 
\rangle \times \left(\langle \hat{\mathbf{S}} 
\rangle \times \boldsymbol{\xi} \right) 
\end{eqnarray}
Similar, the quantum mechanical analog of the stochastic
Landau-Lifshitz can be derived using the same Heisenberg Hamiltonian
together with the Schr\"odinger equation (\ref{TDSE_LL}) or by making
the same transformation we have used to derive the Schr\"odinger
equation (\ref{TDSE_LLG}).   

An alternative way to include temperature, where we assume that the
system is already in equilibrium, is to use statistical operator
\cite{garaninAdvChemPhys11,blumBook}:  
\begin{equation} \label{StatOp}
\hat{\rho}_{\mathrm{Stat.}} =
\frac{e^{-\beta\hat{\mathrm{H}}}}{\mathrm{Tr}\left( e^{-\beta\hat{\mathrm{H}}}
  \right)} \;,
\end{equation}
with $\beta$ the well known inverse temperature $\beta =
\frac{1}{k_BT}$. Then, the time dependence of $\hat{\rho}_{\mathrm{Stat.}}$
is given by:  
\begin{equation} \label{RhoTimeStat}
\hat{\rho}_{\mathrm{Stat.}}(t) = \hat{U}(t)
\hat{\rho}_{\mathrm{Stat.}}(0)\hat{U}^+(t) \;. 
\end{equation}  
$\hat{U}(t)$ is the unitary operator
\begin{equation}
\hat{U}(t) = e^{-i\hat{\mathrm H}t} e^{-\lambda\hat{\mathrm H}t}e^{\lambda
  \langle \hat{\mathrm H} \rangle t} \;, 
\end{equation}
which we have already seen in Eq.~(\ref{PsiTime}). At this point, we
have to notice that Eq.~(\ref{RhoTimeStat}) is a self-consistent equation,
because the unitary operator $\hat{U}(t)$ contains $\langle
\hat{\mathrm H} \rangle$ which has to be calculated with
$\hat{\rho}_{\mathrm{Stat.}}$: $\langle \hat{\mathrm H} \rangle =
\mathrm{Tr}(\hat{\rho}_{\mathrm{Stat.}}\hat{\mathrm H})$. However,
$\hat{\rho}_{\mathrm{Stat.}}$ already needs $\langle \hat{\mathrm H}
\rangle$ to be calculated. Alternatively, we can use (\ref{StatOp}) in
the Liouville equation (\ref{Liouville}) to describe the dynamics. 

\section{SUMMARY} \label{s:summary}
In summary it has been show that the following
non-Hermitian Hamilton operator $\hat{\cal H} = \hat{\mathrm{H}} - i
\lambda\hat{\mathrm{H}}$ inserted in the time dependent Schr\"odinger
equation leads to an equation of motion similar to the classical
Landau-Lifshitz Eq.~(\ref{EOM_LL}). To show this the corresponding
Liouville Eq.~(\ref{Liouville}) as well as Heisenberg
Eq.~(\ref{Heisenberg}) have been derived and discussed. 

It is known that the Landau-Lifshitz equation fails in the huge
damping limit \cite{kikuchiJAP56}. However, the
Landau-Lifshitz-Gilbert (LLG) equation can be simply obtained from the
Landau-Lifshitz equation by rescaling the time. The LLG equation
itself shows the correct physics. The same rescaling has been used to
derive the time dependent Schr\"odinger Eq.~(\ref{TDSE_LLG}) which
corresponds to the classical Landau-Lifshitz-Gilbert
Eq.~(\ref{clLLG}).  

During the complete derivation and later in the manuscript the
similarities and differences of the classical and quantum spin
dynamics have been discussed. It has been shown that only for single 
spins with a Zeeman Hamiltonian $\hat{\mathrm{H}} = -\mathbf{B} \cdot
\hat{\mathbf{S}}$ and in the case of a ferromagnetic multi-spin system
we can expect a classical behavior, which means similar trajectories
of the quantum mechanical expectation values $\langle
\hat{\mathbf{S}}_n \rangle$ and classical spins $\mathbf{S}_n$. (The
index $n$ stands for the $n$th spin.) In all other cases we expect 
a deviation of the trajectories of the expectation values $\langle
\hat{\mathbf{S}}_n \rangle$ from the classical behavior. The reasons
are the appearance of additional terms of the order of $\hbar$ in the
Heisenberg equation which disappear in the classical limit: $S
\rightarrow \infty$ and $\hbar \rightarrow 0$ and in 
the case of a multi-spin system the appearance of entanglement. The
entanglement itself is a pure quantum effect and also disappears in
the classical limit.  

All the results have been discussed theoretical with analytical
calculations and proved with computer simulations. In all cases the
theoretical estimated effects have been seen in the
simulations.   

In the last section of the manuscript two ways have been shown how to
include temperature which has not been taken into account in the
previous sections (Sec.~\ref{s:LLG} - \ref{s:simulations}). The first
way is to add a random noise. This way allows us to investigate the dynamics
under the influence of temperature as well as quantum
fluctuations. The second way deals with the statistical operator. In
this case fluctuations play no role. This way can be used if we
are interested in system which are always in the thermodynamical
equilibrium. This means, non-equilibrium effects cannot be described
with this way.                

\begin{acknowledgments}
The author thanks D. Altwein, M. Krizanac, E. Vedmedenko, Y. Saadi,
M. Maamache, and P.-A. Hervieux for helpful discussions and comments.
This work has been supported by the Deutsche Forschungsgemeinschaft in the
framework of subproject B3 of the SFB 668 and by the Cluster of Excellence
``Nanospintronics''. 
\end{acknowledgments}

\bibliography{Cite}

\end{document}